\documentstyle[11pt,natbib,newpasp,twoside,epsf]{article}
\newcommand{\czw}{^{12}\mathrm{C}}
\newcommand{\cdr}{^{13}\mathrm{C}}
\newcommand{\nvi}{^{14}\mathrm{N}}
\newcommand{\msun}{{\rm M}_\odot}
\newcommand{\spr}{\mbox{$s$-process}}
\def\edcomment#1{\iffalse\marginpar{\raggedright\sl#1\/}\else\relax\fi}
\reversemarginpar

\begin{document}
\title{CNO in Low- and Zero-Metallicity AGB Stars}
 \author{Falk Herwig}
\affil{University of Victoria, 3800 Finnerty Rd, Victoria, BC, Canada}

\begin{abstract}
The oldest stars hide information about the
chemical evolution of the early universe. 
In this contect new models of the AGB stellar evolution phase at very low
and zero metallicity are presented.
Due to the deficiency or absence of CNO catalytic
material hydrogen burning operates at significantly higher
temperatures than in stars of higher metallicity. As a result
convective mixing plays a very important role since the nuclear
burning shells are not well separated by an entropy barrier. These
stars can host flash-like burning events induced by mixing of $\czw$
and protons at the core-envelope interface. Three different mixing events with
proton-capture nucleosynthesis can be encountered. One of
them, the hot dredge-up, is reported here for the first time. All
these specific 
modes of nucleosynthesis are relevant for the production of CNO
material in the first intermediate mass stars. 
\end{abstract}

\section{Introduction}
The first nuclear processing of pristine Big Bang matter has left a
signature in today's most metal poor low mass stars. 
There is some debate whether both low mass and
massive stars contributed to this processing (Nakamura \& Umemura, 2002) or
whether the first stars were exclusively very massive (e.g.\ Abel et
al., 2002). The recently 
discovered low mass star \mbox{HE0107-5240} has [Fe/H]=-5.3
(Christlieb et al., 2002). This may indicate that the  Pop.\,III
IMF has a low mass component. The large
nitrogen overabundance in this object could mean that intermediate
mass stars are involved in the contamination of this star.

This is much more obvious for stars like \mbox{HE0024-2523}
(Lucatello et al., 2003), a very metal poor main sequence star with
[Fe/H]=-2.7, an \spr\ abundance signature, large C and N
overabundances as well as radial velocity variations. It has been
suggested that this star and related objects have been polluted by an
intermediate mass binary companion.
The carbon and \spr\ abundance could -- in principle --
be understood with the dredge-up and \spr\ mechanisms known from AGB
stars of solar-like metallicity (e.g.\ Gallino et al., 1998). 
The nitrogen  
overabundance  could result from partial envelope CNO cycling (hot
bottom burning, HBB). Siess et al.\ (2002) predict that both carbon as 
well as nitrogen may be significantly enhanced due to dredge-up and HBB
in Z=0 AGB stars. Their $3(2)\msun$ model predicts at the last
thermal pulse (TP)  [C/H]$\sim -0.9(0.0)$, [N/H]$\sim +0.4(-2.2)$ and
C/O$=5(53)$. \mbox{HE0024-2523} has  [C/H]$= -0.1$, [N/H]$=
-0.6$ and C/O$=100$. A [C/N] ratio of a few tenth of a dex
is also found in other similar stars (e.g.\ \mbox{CS 22948}, Aoki et al.,
2002). It is even more extreme in \mbox{HE0107-5240}. This puts some 
limit on the role HBB can play because efficient HBB may too quickly
decrease the C/N ratio below unity.
\begin{figure}
\plotone{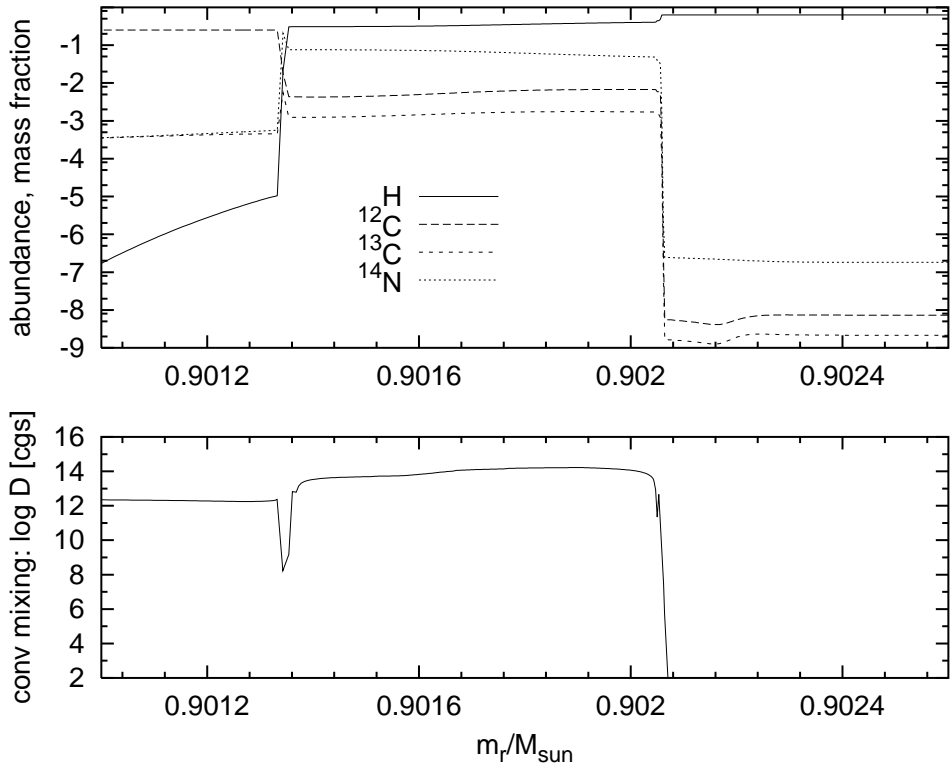}
\caption{\label{fig:FF08348}
Proton ingestion and burning phase during the 10$^{\rm th}$ TP of a
Z=0 sequence with M=$5\msun$.
Top panel: Abundance profile of upper He-flash convection
zone (left), H-flash convection zone (middle) induced by proton
ingestion from above, and H-shell ashes on top of the core
(right). Bottom panel: 
Convective mixing efficiency in terms of a diffusion coefficient.}
\end{figure}

The role of intermediate mass stars as a source of primary
nitrogen in the early chemical evolution has
recently been revisited on the basis of the N abundances in
Damped Lyman-$\alpha$ Absorbers (Pettini et al., 2002, and Henry \&
Proschaska, this volume). Meynet \& Maeder (2002) have
shown that in intermediate mass stars of very low metallicity
rotationally induced mixing can lead to significant $\nvi$ enrichment
of the outer layers even before the first TP occurs.
In these models the C/N ratio is much smaller than unity and
would not reproduce the abundance patterns of stars like
\mbox{HE0024-2523}. Clearly, the  the envelope abundances are modified
in the subsequent TP AGB phase.

In  very low and zero metallicity stars the convection zone driven by the
He-flash may reach into the proton rich unprocessed outer layers . As
protons enter the He-shell environment they initiate a vigorous
non-equilibrium nuclear burning, mainly of proton captures on $\czw$. 
This may happen in the case of the He-core flash as well as the AGB
He-shell flash 
for low enough metallicity and core mass (Fujimoto et al., 2000). The
studies by Siess et al.\ (2002) and Chieffi et al. (2001) found that the TP AGB
evolution of Z=0 objects 
eventually resembles  that of only mildly metal poor AGB stars. Either
the second dredge-up or a 
convective instability of the H-shell at reignition after a TP
sufficiently increases the catalyst material for
the CNO cycle. 

In this paper I present first results on
stellar evolution calculations that consider convective
nucleosynthesis in a numerically consistent way and also allow for
soft convective boundaries with respect to mixing. Models
of the proton ingestion induced H-flash during TPs of Z=0
AGB stars of $2\msun$ and $5\msun$ are described. In addition I 
report on a new 
process of carbon and nitrogen enrichment uniquely present in very low
metallicity stars which is a combination of HBB and
third dredge-up. 
\begin{figure}
\plotone{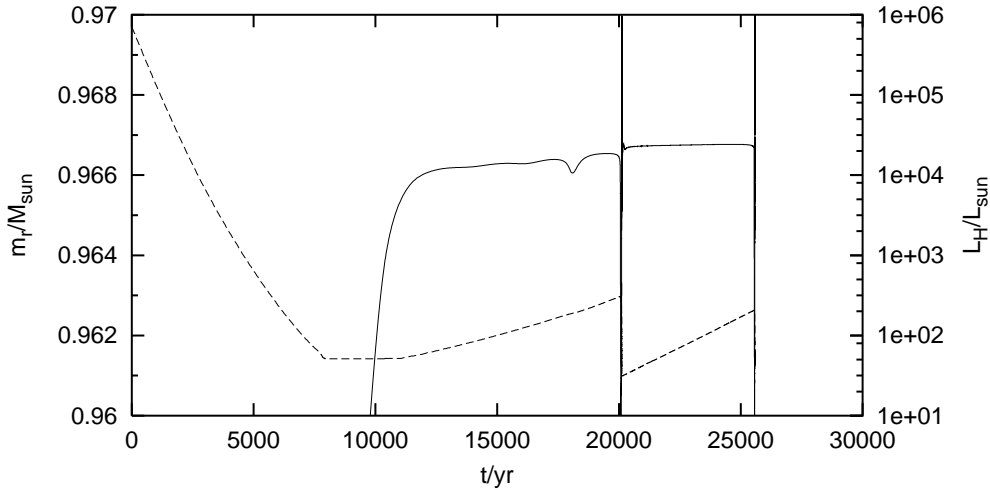}
\caption{\label{fig:MH-LH-M5Z0.0001}
Evolution of the mass coordinate of the H-free core $M_{\rm H}$ and
the H-burning luminosity $L_{\rm H}$ during the end of the second
dredge-up ($t<7500{\rm yr}$) and the first two AGB TPs
of the Z=0.0001 sequence with M=$5\msun$.
}\end{figure}

\section{Stellar evolution code and physical model}
The 1D hydrostatic stellar evolution code EVOL has been previously
used for extensive calculations of the TP AGB phase
(Herwig, 2000). The models contain a parametric
hydrodynamic overshooting
at all convective boundaries. The efficiency $f$ is expressed as the
e-folding distance of the decay of the turbulent velocity field in the
stable layer. This scheme has been adopted from  2D RHD simulations.
In AGB models the efficiency can be constrained
by carefull analysis of the \spr\ nucleosynthesis, and comparison with
both stellar observations and measurements of pre-solar meteoritic
grains. In particular 
certain temperature dependent branchings provide valuable constraints
on mixing processes in low mass TP AGB stars
(Lugaro et al., 2002; Herwig et al., 2002).

EVOL has also been used to construct model sequences of
pre-white dwarf He-shell flash stars which evolve into born-again AGB
stars (Herwig et al., 1999). In these stars the convection zone of the
He-shell flash can 
reach into the thin H-rich envelope, very similar to the events
reported by Fujimoto et al.\ (2000) in very low or zero metallicity
stars. We follow the convective nucleosynthesis 
with a solution scheme in which nuclear rate equations
 and time dependent mixing equations  are fully
coupled. In that way the abundance profiles, in particular of
hydrogen, reflect at each grid point and time step the simultaneous
action of fast convective mixing and fast p-captures by
$\czw$. Although the  numerical scheme is treating the abundance
evolution in this convective nucleosynthesis event correctly
there is evidence that the physical model to describe convection (here
the mixing length theory) may need to be adjusted
(Herwig, 2001), which is done here in an approximated way.
\begin{figure}
\plotone{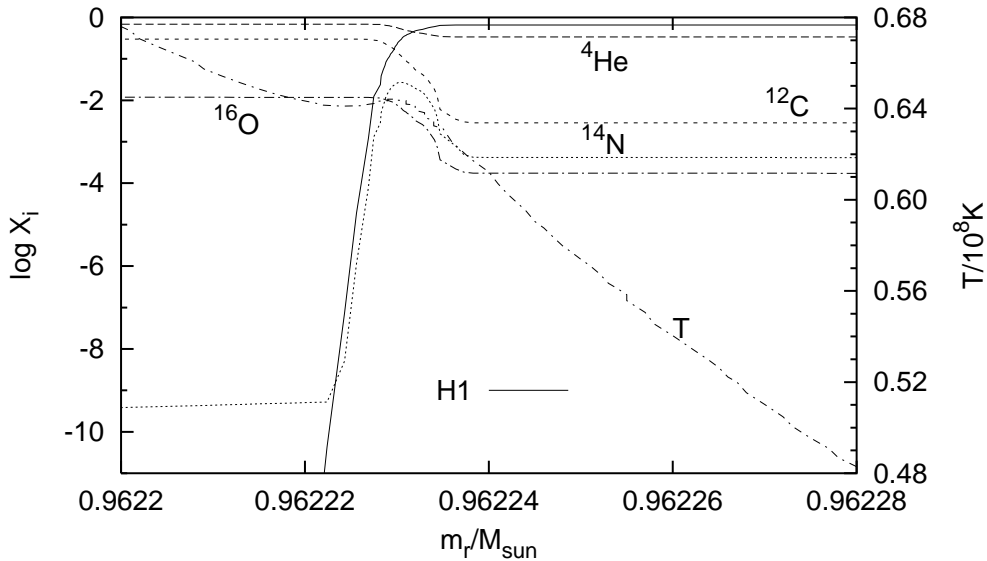}
\caption{\label{fig:Z-4_TP1_HDUP}
Temperature and abundance profile during the \emph{hot dredge-up} phase after
the first TP of the $5\msun$, Z=0.0001 evolution sequence.
}
\end{figure}

\section{Proton capture burning in the He-shell flash layer}
I have computed Z=0 tracks for initial stellar mass of $2\msun$ and
$5\msun$ from the pre-main sequence up to the TP AGB. In both cases
protons are eventually entering the He-flash convection
zone. The $5\msun$ sequence starts on the AGB with eight weak
TPs with $L_{\rm He}<10^{5}{\rm L_\odot}$ and a TP period of $\approx 3500{\rm
yr}$ . The H-shell becomes convectively unstable at the time it
reignites after the $9^{\rm th}$ TP. This occurence has been
previously reported by Chieffi et al.\ (2001) and Siess et al.\ (2002) as a
\emph{hydrogen convective episode} (HCE). The trace of any small C and
N envelope abundance increase resulting from this event is lost in the
following TP. As the He-flash convection zone develops it
eventually reaches out into the proton rich envelope and the well know
sequence of events, including the H-flash and the splitting of the
convection zone take place. A snapshot of the developing H-flash
convection zone on top of the He-flash convection zone is shown in
Fig.\,\ref{fig:FF08348}. In comparison to the very late TP
in some pre-white dwarf models (Herwig et al., 1999)  much more 
hydrogen is available and a large amount of $\nvi$
is produced. The processed material is dredged-up to the surface in two
subsequent phases associated with the H-flash and the He-flash
respectively. The envelope $\nvi$ mass fraction  after the event is
$2.5 \cdot 10^{-5}$. In the $2\msun$ case the proton ingestion into
the He-shell flash layer happens after 7 weak initial pulses without a
previous HCE event. A peak H-luminosity of almost $10^{10}{\rm
L_\odot}$ is obtained and after the event both the $\czw$ and the
$\nvi$ abundance in the envelope amounts to about $10^{-4}$ in mass fraction
with C/N$<1$. For both masses the further evolution is equivalent to
models at higher Z with the same supply of CNO catalytic material. 

If the models by Meynet \& Maeder apply to  Z=0 intermediate mass
stars, and if these are rapid rotators, then the proton ingestion into
the He-flash convection layer will not occur in TP AGB models. It is
also unlikely that the HCE will occur in this case.
\begin{table}[t]
\caption{\label{tab:abund}Surface abundances before and after the first TP
(sequence $5\msun$, Z=0.0001) sequence with  hot dredge-up}
\begin{tabular}{rrrrrrr}
\hline
 & X($\czw$)&X($\cdr$)&X($\nvi$)&$\czw/\cdr$&$\czw/\nvi$&$\cdr/\nvi$ \\
\hline
before TP& 6.7E-6&3.6E-7& 2.2E-5& 16& 0.26&0.016 \\
after TP & 9.5E-5& 1.6E-5& 5.2E-5& 6 & 1.8& 0.31 \\ 
\hline
\end{tabular}
\end{table}

\section{Hot dredge-up}
In addition to the Z=0.0 sequences I  calculated tracks for $\log Z=-5$,
$-4$, and $-3$ for $2\msun$ and $5\msun$. In all sequences
efficient dredge-up is present leading to gradual CNO enrichment at
each TP. A new phenomenon was encountered in all $5\msun$
cases (Fig.\,\ref{fig:MH-LH-M5Z0.0001}). During the dredge-up phase
protons are partially mixed with the 
underlying $\czw$ in the core. In this tiny layer the H/$\czw$ ratio
drops by orders of magnitude. At the given temperature this leads to
 fast proton captures and a very large H-burning luminosity.
The partial mixing layer in these calculations is the immediate
consequence of the hydrodynamic overshooting model.  
This effect is only observed  in very low Z models where the
dredge-up layer is hotter  compared to higher metallicity models. As
shown in Fig.\,\ref{fig:Z-4_TP1_HDUP} a substantial $\nvi$ pocket
forms in this partial mixing layer. A $\cdr$ pocket of the same size
(not shown) is also present. The \spr\ can not take place in this
partial mixing layer because the time scale is too short and the
$\nvi$ abundance is too high. In these models carbon is CN processed
\emph{on the fly} as it is dredged-up to the envelope. This 
process can be thought of as a combination of HBB and
third dredge-up, and I call it \emph{hot dredge-up}. The dredge-up with
simultaneous hydrogen burning may be much deeper than usually
found. This can boost the envelope enrichment with processed matter,
in particular C and N. We summarize the C and N abundance features
before and after the first thermal pulse 
with hot dredge-up in Table\,\ref{tab:abund}. It is tempting to compare
these results  with observed C and N abundances and isotopic ratios, for
example Tab.\,6 in Aoki et al.\ (2002). The isotopic ratios agree within a
factor of two. However, the combined effect of hot dredge-up, HBB, and
mass loss over many thermal pulses should be taken into
account before 
detailed comparisons are useful.

\section{Conclusions}
All recent calculations of the  AGB phase at Z=0 agree that thermal
pulses and dredge-up occur. Various processes, in particular
rotationally induced mixing during the He-core burning phase, can lift
the envelope 
abundance of catalytic material for the CNO cycle so that eventually
the evolution in these objects resemble those of only mild metal
deficiency. It is therefore not clear if the proton ingestion into the
He-shell flash convection zone or the HCE take place in real Z=0 AGB
stars should they have existed. 
Here a hot variant of the third dredge-up was described which may be
an important ingredient in understanding both the primary production
of nitrogen as well as the peculiar carbon and nitrogen overabundances
observed in very metal poor stars.

\paragraph{Acknowledgments:} I would like to thank D.A. VandenBerg
for support through his Operating Grant from the Natural Science and
Engineering Research Council of Canada.

\end{document}